\documentclass[preprint,review, 11pt]{elsarticle}
\usepackage[margin=1in]{geometry}

\usepackage{graphicx}
\usepackage{float}
\usepackage{ctable}
\usepackage{algorithm}
\usepackage{wrapfig}
\usepackage{algpseudocode}
\usepackage{multirow}
\usepackage{amsfonts}
\usepackage{amsmath}
\usepackage{booktabs}
\usepackage{makecell}
\usepackage[ruled,vlined,linesnumbered,algo2e]{algorithm2e}
\usepackage{array}
\graphicspath{{figs/}}

\newcommand{\vect}[1]{\boldsymbol{#1}}
\graphicspath{{figs/}}

\journal{Computers in Biology and Medicine}

\graphicspath{{figs/}}

\begin{document}
	
	\begin{frontmatter}

			\title{Automated Identification of Atrial Fibrillation from Single-lead ECGs Using Multi-branching ResNet}
			\author{Jianxin Xie}
			\author{Stavros Stavrakis}
			\author{Bing Yao\corref{cor1}}
			
			\cortext[cor1]{Corresponding author: byao3@utk.edu; \\Jianxin Xie and Bing Yao are with the Department of Industrial \& Systems Engineering, The University of Tennessee, Knoxville, TN, 37996 USA.\\Stavros Stavrakis is with University of Oklahoma Health Sciences Center, Oklahoma City, OK 73104 USA.}

	\begin{abstract}
     Atrial fibrillation (AF) is the most common cardiac arrhythmia, which is clinically identified with irregular and rapid heartbeat rhythm. AF puts a patient at risk of forming blood clots, which can eventually lead to heart failure, stroke, or even sudden death. It is of critical importance to develop an advanced analytical model that can effectively interpret the electrocardiography (ECG) signals and provide decision support for accurate AF diagnostics. In this paper, we propose an innovative deep learning method for automated AF identification from single-lead ECGs. We first engage the continuous wavelet transform (CWT) to extract time-frequency features from ECG signals, Then, we develop a convolutional neural network (CNN) structure that incorporates ResNet for effective network training and multi-branching architectures for addressing the imbalanced data issue to process the 2D time-frequency features for AF classification. We evaluate the proposed methodology using two real-world ECG databases. The experimental results show a superior performance of our method compared with traditional deep learning models.
	\end{abstract}


\end{frontmatter}

\section{Introduction} \label{s:intro}
 
 %{\color{red}It's still not in IEEE citation format}

Cardiovascular diseases have been the leading cause of mortality globally. The World Health Organization (WHO) states that about 17.9 million people perish due to cardiovascular disease each year, contributing 32\% to the worldwide death toll. Atrial fibrillation (AF) is the most common cardiac arrhythmia caused by uncoordinated electrical activities in the atria. Although AF itself does not lead to a lethal condition, it will substantially increase the risk of catastrophic diseases such as heart failure, stroke, and sudden death \cite{lubitz2013atrial,bernstein2021effect}. The prevalence of AF plagues over 2.7 million people in the United States, and this number is estimated to rise to 12.1 million in 2030, as the population ages \cite{colilla2013estimates}. %Many factors may potentially engender AF, such as coronary artery disease, hypertrophic cardiomyopathy, obstructive sleep apnea, obesity, and hyperthyroidism \cite{weiss2011subclinical, chandra2017atrial,goldberger2015evaluating}. 
In healthcare practice, the electrocardiogram (ECG) is a cost-effective and noninvasive medical approach to record the electrical signals on the body surface as a reflection of cardiac health conditions \cite{yao2020spatiotemporal,yao2021constrained,yao2016physics}. %Despite ECG is a well-developed sensing technique, the identification of AF still stands as a challenging task: 

%\begin{itemize}
%\item  The current diagnosis nowadays greatly depends on emergence of specific symptoms, such as paltipation, chest discomfort, dyspnea, and dizziness \cite{xia2018detecting}. However, the occurance of AF is asymptomatic and may appear briefly in ECG records \cite{savelieva2000silent, ghiasi2017atrial}. Besides, the AF-like symptoms may not necessarily appear on patients who actually contract AF, meanwhile up to 40\% of atrial fibrillation (AF) patients are asymptomatic \cite{xiong2015asymptomatic}. % This significantly escalates the difficulty for precision diagnosis of AF.
%\item In clinical practice, the ECG signals are generally examined by well-trained medical doctors to search for the abnormalities based on visual evaluation and manual interpration. The non-stationary and non-linear properties of ECG singnals, and other uncertain factors make the manual AF detection a very challenging and time consuming task \cite{jing2021ecg}. Moreover, under circumstances where no cardiac experts are available, such as in primary care center or emergecy units, the automation identification of underlying cardiovascular disease is of paramount importance. 
%\end{itemize}

%Therefore, an effective automatic detector is urgently needed for AF diagnosis. 

A wide range of data-driven approaches have been developed to investigate ECG signals for AF detection. Most traditional data-driven methods focus on extracting morphological features and heart rate variability from ECG waveform signals to detect AF, which depends heavily on manual feature engineering. Deep Neural Network (DNN), which does not require explicit feature engineering, is another powerful tool that has achieved promising results in data-driven disease detection. Various DNN models including convolutional and recurrent neural networks (i.e., CNNs, RNNs) have been designed for AF detection and outperformed conventional machine learning methods. Despite many advances in data-driven ECG analysis, there are three major challenges that remained to be tackled: (1) The ECG recordings collected from clinics are often in PDF format. An effective preprocessing procedure is needed to retrieve the raw ECG signals from the PDFs before being fed to the machine learning models. (2) ECG signals are generally composed of multiple frequency components. The DNN models built upon raw ECG time series may not fully exploit the time-frequency information inherent in the ECG signals. (3) Note that the learning capacity for a DNN often increases when the network goes deeper. However, the deeper structure can result in gradient dissipation problems, leading to unsatisfactory prediction performance. (4) Data-driven AF detection also suffers from the common issue of imbalanced data in machine learning (e.g., AF ECG samples are much less compared to normal ECG samples). The classifier directly built from the imbalanced data will generate biased and inaccurate predictions.

In this paper, we develop an automatic AF detector based on continuous wavelet transform (CWT) and 18-layer Residual Neural Network (ResNet18) with a multi-branching structure (CWT-MB-ResNet). We first develop a preprocessing procedure to extract ECG signals from ECG PDFs and leverage the CWT to transform the extracted signals into the time-frequency domain. Second, ResNet18 is engaged to alleviate the gradient dissipation problem in deep-structured networks, allowing it to learn deeper features from 2D time-frequency image data and achieve better performance. Finally, we propose to incorporate a multi-branching output structure adapted from our prior work \cite{wang2021multi} into the ResNet to deal with the issue induced by the imbalanced dataset in AF identification. 
The multi-branching technique exempts artificial data augmentation and does not require any preassumptions. 
The performance of the proposed framework is evaluated by two real-world datasets: PhysioNet/CinC challenge 2017 \cite{goldberger2000physiobank,clifford2017af} and ECG data obtained from the University of Oklahoma Health Sciences Center (OUHSC). 
Experimental results show that our CWT-MB-ResNet significantly outperforms existing methods commonly used in current practice.

The rest of this paper is organized as follows: Section \ref{s:background} presents the literature review of existing data-driven methods for AF detection. Section \ref{s:methods} introduced the data processing details and the proposed prediction method. Section \ref{s:results} shows the experimental results in AF identification. Section \ref{s:conclusion} concludes the present investigation.

\section{Research Background}
 \label{s:background}
Traditional machine learning approaches focus on the extraction of ECG morphological features \cite{de2004automatic} and heart rate variability information \cite{park2009atrial} to identify AF conditions. Those methods are mostly in light of two aspects of AF-altered ECG characteristics: (1) the absence of P wave or fibrillatory P waves presented as oscillations in low amplitude around the baseline \cite{ladavich2015rate}; (2) irregular R-R intervals \cite{oster2015impact}. %{\color{red} Are the following methods based on the aforementioned 2 aspects?yes} 
Multiple feature-based automation techniques have been proposed to classify AF-altered ECGs, such as linear discriminant analysis \cite{de2004automatic}, support vector machine \cite{islam2017atrial, billeci2017detection}, independent component analysis\cite{ye2012heartbeat}. When there exists a high level of noise or faulty detection, the performance of AF identification methods that solely study the P wave deteriorates significantly due to the chaotic signal baseline introduced by the noise \cite{larburu2011comparative}. Most R-R interval-based methods \cite{lian2011simple, tateno2001automatic} usually require long ECG segments to detect AF episodes, and become ineffective when it comes to short ECG signals (less than 60s) or in the presence of significant sinus arrhythmia or frequent premature atrial contractions \cite{xia2018detecting}. Moreover, traditional methods require a separate feature extraction process before feeding the data into the classifier, as well as manually establishing the detection rules and threshold. This can be computationally expensive and may not generalize well when applied to a larger population. 

In the past few decades, deep learning or deep neural network (DNN) has emerged as a powerful tool for pattern recognition that can learn the abstracted features from complex data and yield state-of-the-art predictions \cite{mousavi2019ecgnet, xie2022physics,chen2022prediction,wang2022multi,xie2022physics1}. As opposed to traditional machine learning, deep learning presents strong robustness and fault tolerance to uncertain factors, which makes it suitable for beat and rhythm classification from ECG data \cite{tutuko2021afibnet}. Moreover, existing research has indicated that deep learning methods demonstrate more efficient and more potent predictive power than classical machine learning methods for AF identification \cite{murat2021review, cai2020accurate}. Various neural network structures have been proposed to address the heart disease detection problems such as recursive neural networks (RNNs) \cite{singh2018classification}, long short-term memory (LSTMs) \cite{yildirim2018novel}, and autoencoders (AEs) \cite{nurmaini2020deep}. Among them, CNN models have wide applications in AF detection.  CNNs commonly excel at 2D data processing such as image-based classification. Recent literature has also shown that 1D-CNNs generate superior predictive results compared to traditional DNNs and RNNs when processing 1D ECG signals \cite{xiong2017robust, tutuko2021afibnet, wang2023hierarchical}. 1D-CNN stands as an effective tool to extract the morphological features and learn the slit temporal variations in time series data \cite{tutuko2021afibnet,nurmaini2020robust}.  However, existing literature \cite{ullah2021hybrid,wu2018comparison} has demonstrated that the predictive accuracies produced by 1D CNNs are lower than their structure-alike 2D counterpart in ECG classification due to more comprehensive information in the 2D input data and the more superior capability of 2D CNN in feature extraction and interpretation.

Owing to its outstanding performance and strong ability in pattern recognition, 2D CNN has been explored for ECG classification by virtue of its capacity to smartly suppress measurement noises and extract pertinent feature maps using convolutional and pooling layers \cite{huang2019ecg}. For example, Izci et al. \cite{izci2019cardiac} engaged a 2D CNN model to investigate ECG signals for arrhythmia identification. They segmented the ECG signals into heartbeats and directly converted each heartbeat into grayscale images, which served as the input of the 2D CNN model. Similarly, Jun et al. \cite{jun2018ecg} proposed to combine 2D CNN and data augmentation with different image cropping techniques to classify 2D grayscale images of ECG beats. However, these end-to-end 2D CNNs are directly fed with original ECG beat segments without considering the possible noise contamination. Moreover, the 2D input data were created by directly plotting each ECG beat as a grayscale image with unavoided redundant information residing in the image background. This procedure requires extra storage space for training data and increases the computational burden without extracting critical features inherent in the ECG beats.

ECG signals generally consist of various frequency components, which can be used to identify disease-altered cardiac conditions. Wavelet transform (WT) \cite{daubechies1990wavelet,van2019cost,yao2017characterizing} has been proven to be a useful technique for extracting critical time-frequency information pertinent to disease-altered ECG patterns \cite{kutlu2012feature, he2018automatic}. As such, WT is favored as a feature-preprocessing procedure that converts 1D ECG signals into 2D images containing time-frequency features. The resulting 2D feature images then serve as the input of CNNs for ECG classification instead of the original 2D ECG plots. 
For instance, Xia et al. \cite{xia2018detecting} engaged the short-term Fourier transform (STFT) and stationary wavelet transform to convert ECG segments into 2D matrices which were then fed into a three-layer CNN for AF detection. Wang et al. \cite{wang2021automatic} combined the time-frequency features extracted by Continuous Wavelet Transform (CWT) and R-interval features to train a 2D CNN model for ECG signal classification. Wu et al.  \cite{wu2019novel} built a 2D CNN based on time-frequency features of short-time single-lead ECGs extracted from three methods, i.e., STFT, CWT, and pseudo Wigner-Ville distribution, to detect arrhythmias. Huang et al. \cite{huang2019ecg} developed an ECG classification model by transforming ECG signals into time-frequency spectrograms using STFT and feeding them into a three-layer 2D CNN. Li et al. \cite{li2019ventricular} included three different types of wavelet transform (i.e., Morlet wavelet, Paul wavelet, Gaussian Derivative) to create 2D time-frequency images as the input data to the 2D CNN-based ECG classifier. 

In addition to effective information extraction from ECG time series, the realization of the complete data potential is heavily reliant on advanced analytical models. Although the abovementioned works have validated the superiority of 2D CNN-based approaches, the shallow network structures with a limited number of layers can potentially hinder the extraction of deeper features. Naturally, the capacity for a neural network to learn is enhanced by an increase in the number of layers. However, having a deeper network structure can result in the gradient dissipation problem, which impedes convergence during network training, leading to suboptimal prediction performance. To cope with this issue, the residual neural network (ResNet) has been developed with an important modification, i.e., identity mapping, induced by the skip connection technique \cite{he2016deep}, which has wide applications in classifying the ECG signals. For example, Jing et al. \cite{jing2021ecg} developed an improved ResNet with 18 layers for single heartbeat classification. Park et al. \cite{park2022study} used a squeeze-and-excitation ResNet with 152 layers and compared the model performance trained by ECGs from a 12-lead ECG system and single-lead ECG data. Guan et al. \cite{guan2022ha} proposed a hidden attention ResNet to capture the deep spatiotemporal features using 2D images converted from ECG signals.

Automated ECG classification also suffers from the long-standing issue of imbalanced data in machine learning. Diverse sampling and synthetic strategies have been proposed to address the imbalanced data issue, which focus on creating a balanced training dataset out from the original imbalanced data to mitigate the potential bias introduced by imbalanced data distribution during model training \cite{he2009learning}. Frequently employed techniques consist of random over-sampling and under-sampling, informed adaptive undersampling, and synthetic minority over-sampling technique (SMOTE) \cite{gao2019effective, wang2021multi, qiu2022optimal}. For example, Luo et al. \cite{luo2021multi} engaged SMOTE to synthesize minority samples and create a balanced training dataset for automated arrhythmia classification. Ramaraj et al. \cite{ramaraj2021novel} incorporated an adaptive synthetic sampling process into the training of deep learning models built with gated recurrent units to address the class imbalance problem for ECG pattern recognition. Nurmaini et al. \cite{nurmaini2020robust} compared sampling schemes of SMOTE and random oversampling with RNN and concluded that the balanced dataset created by SMOTE significantly improved the classification performance. In addition to fabricating balanced ECG datasets, Gao et al. \cite{gao2019effective} and Petmezas et al.  \cite{petmezas2021automated} proposed to engage dynamically-scaled focal loss function to suppress the weight of loss corresponding to the majority class, so that their contribution to the total loss is reduced to alleviate the class imbalance problem. However, this method requires the preassumption of a focusing parameter to modulate the effect of the majority class on the total loss. Existing methods mainly focus on using sampling and synthetic strategies or modifying the loss function, little has been done to create new network structures without making extra assumptions and feature engineering to cope with the imbalanced data issue in AF identification from ECG signals.

\section{Materials and Methods} \label{s:methods}

\subsection{Dataset} 
In this study, two AF databases from different sources, i.e., ECG recording from PhysioNet/CinC challenge 2017 and ECG PDFs from OUHSC, are used to evaluate the performance of data-driven detection methods. Both databases consist of short single-lead ECG recordings for AF and non-AF patients. PhysioNet/CinC Challenge 2017 is an open database including 8528 single-lead ECG signals and their annotations. Among them, 5050 ECG recordings are labeled as normal sinus rhythm while 738 signals are annotated as AF. The sampling frequency of recordings is 300 Hz and the duration of ECG signals varies from 9s to 30s. The OUHSC database contains ECG signals in PDF format with 33 recordings from AF subjects and 227 normal samples, which are annotated by cardiologists from OUHSC. Each recording has a duration about 30s with a sampling frequency of 60 Hz. We use 80\% of the total data for training and the remaining 20\% for testing for both databases.

\subsection{ECG Signal Preprocessing}
  
\begin{figure}
	\centering
	\includegraphics[width=2.8in]{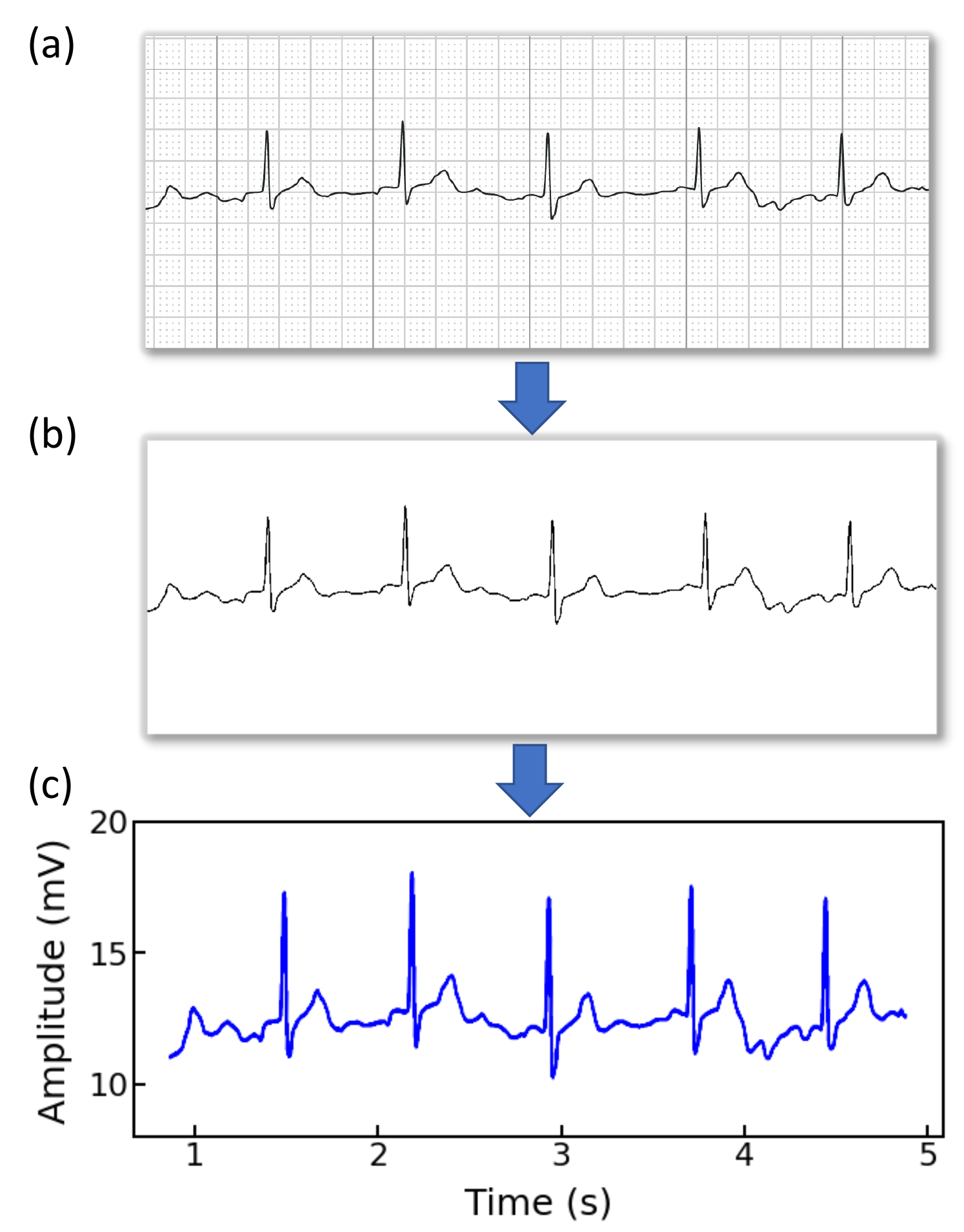}
	\caption{An example of (a) a raw image recording of an ECG segment in PDF format, (b) the ECG image that filters out the grid background, (c) the digitalized ECG time series signal.
	}
	\label{Fig:oupdf}
\end{figure}

  Note that the original ECG recordings from OUHSC are in PDF format, as shown in Fig. \ref{Fig:oupdf}(a). It is necessary to accurately extract the numerical ECG readings from the PDF files for further data preprocessing and analysis, which is achieved by the following procedure:
  \begin{itemize}
  	\item \textit{Transforming PDF files into gray-scale images represented by 2D pixel matrices}: We discretize the 2D image into a pixel matrix. Then, each pixel is converted to a fixed number of bits to represent the gray-scale intensity of the corresponding point in the image. As shown in Fig. \ref{Fig:oupdf}(a), the ECG signals are displayed in the darkest color on the plot with the color intensity of 1, i.e., $h(m,n) = 1$, while the grid lines appear in a lighter color, i.e., $0<h (m,n)< 1$, where $h(m,n)$ denotes the color intensity of the pixel at column $m$ and row $n$. Note that the background color intensity is 0. 
  	\item \textit{Removing grid lines from the ECG plot}: We replace the pixel shade values of the grid lines with the background color value: i.e., $h(m,n \mid h(m,n)<1) = 0$. This allows the ECG signals to distinguishably stand out, as illustrated in Fig. \ref{Fig:oupdf}(b). The quantized image is thus encoded into a binary digital format, i.e., black as "1" and white as "0". As such, the entire ECG image is transformed into a binary digital matrix without the grid lines.
  	\item \textit{Extracting the digital ECG time series}: The positions of the black pixels (i.e., ECG signal) in the binary matrix are further extracted, which are represented mathematically as a set of $(m, n)$ pairs:
  	$$S =\{(m, n) | h(m, n) = 1\}$$
  	The resulted $S$ is then used to reconstruct the digital ECG time series, where $m$ stands for the time course, and $n$ correspond to the magnitude of the ECG signal. As such, we are able to extract the ECG recordings from the PDFs to digitalized ECG time series signals (Fig. \ref{Fig:oupdf}(c)), which will be used for further processing and model training. 
  \end{itemize}

  %{\color{red} (Rewrite this part in a clearer way. Plot a flowchart to illustrate each step. No need to mention the concatenation). First, the PDF files are transfromed to images. These images are then converted from true color image RGB to grayscale images for further information extraction. The shade of each pixel on the grayscale image can be represented by a number, such that the whole ECG image can be interpereted and processed as a matrix where the value of each element denotes a color scale and the numbers of the row and the column of the element reflect the location on the ECG image. The ECG signals are ploted in the darkest color whereas the gridlines are shown in a lighter shade. Based on that, we can easily eliminate the gridlines from the ECG plot by replacing the pixel shade value of the grids with the background color value. Another challenge is that the ECG signals are continous but ploted in 4 lines. The numerical matrix corresbonding to the image is saperated into 4 sub-matrices in coordinate with these 4 lines. We concatenate the signals by assuming that the amplitude of end point of the ECG at current line is equal to that of the starting point of the following line. The amplitude is defined as the row index of one valid ECG pixel in the sub-matrix. The obtained ECG magnitude is normalized according to the vertical length of the image, i.e., the total number of rows in the sub-matrix}. 

Raw ECG recordings are often contaminated by noises, such as baseline wandering, electromyography disturbance, and power-line interference \cite{mian2020baseline}, which will negatively impact the information extraction and model performance. In this work, we engage BioSPPy, a toolbox for biosignal processing 
%({\color{red} include what kinds of preprocessing steps? More details are needed.}) 
written in Python, for ECG signal denoising. The BioSPPy library provides comprehensive functions for processing ECG signals including functions for importing ECGs, filtering out interfering components, and correcting baseline wandering \cite{biosppy}. Specifically, after loading the ECG data, a high-pass filter is applied to remove the low-frequency noise (e.g., baseline wandering), a notch filter to remove power-line interference, and a low-pass filter to filter out the high-frequency noise.

\subsection{Continuous Wavelet Transform}
ECG signals encompass multiple feature components in both the time and frequency domains. In this study, we engage the continuous wavelet transform (CWT) to extract time-frequency features from ECGs due to its excellent performance in the analysis of transient and non-stationary time series signals \cite{keissar2009coherence}. CWT is the most popular tool for time-frequency analysis that reflects the frequency components of data changing with time. CWT is verified to outperform the traditional STFT due to its ability to provide multi-resolution decompositions of the signal, which allows for a trade-off between time and frequency resolution, i.e., higher frequency resolution for signals with sharp transients and higher time resolution for signals with slow-varying frequency content \cite{dokur2001ecg}. 
 Additionally, compared to discrete wavelet transform (DWT), CWT remedies non-stationarity and coarse time-frequency resolution defects and supports the extraction of arbitrarily high-resolution features in the time-frequency domain \cite{addison2005wavelet}.

The CWT of the ECG time-series signal denoted as $x(t)$ is achieved according to:
\begin{equation}
	T(a, b)=\frac{1}{\sqrt{a}} \int_{-\infty}^{+\infty} x(t) \psi \left(\frac{t-b}{a}\right) \mathrm{d} t
\end{equation}
where $T(a, b)$ stands for the intensity of transformed signals, $\psi(\cdot)$ is the wavelet basis (also known as the mother wavelet), $a$ is the scale factor quantifying the compressed or stretched degree
of a wavelet, and $b$ is the time shift parameter defining the location of the wavelet. %{(\color{red} what is x(t) and what is T(a,b)?)} {\color{red}
The scale can be used to derive the characteristic frequency of the wavelet as \cite{wu2019novel}:
\begin{equation}
	F = \frac{F_c\times f_s}{a}
\end{equation}
where $F_c$ is the center frequency of the mother wavelet
and $f_s $ is the sampling frequency of the signal. This relationship shows that smaller (larger) values of $a$ correspond to higher (lower) frequency components. 
In CWT,  the mother wavelet plays a critical role in time-frequency analysis, the choice of which depends on its similarity with the original signal \cite{ngui2013wavelet}. Here, the Mexican hat wavelet (mexh) is chosen to serve as the mother wavelet because its shape is similar to the QRS waves and it is commonly used in ECG signal analysis \cite{wang2021automatic}. Specifically, the mexh is the second derivative of a Gaussian function \cite{addison2005wavelet}, which is defined as:
\begin{equation}
	\psi(t)=\frac{2}{\sqrt{3} \sqrt[4]{\pi}} \exp \left(-\frac{t^2}{2}\right)\left(1-t^2\right)
\end{equation}
Continuously changing the scale factor $a$ and time shift parameter $b$ generates the 2D wavelet coefficients $T(a,b)$, which can be viewed as a 2D scalogram of the ECG signal in both the time and frequency domain \cite{wang2021automatic}.

\begin{figure}
	\centering
	\includegraphics[width=5in]{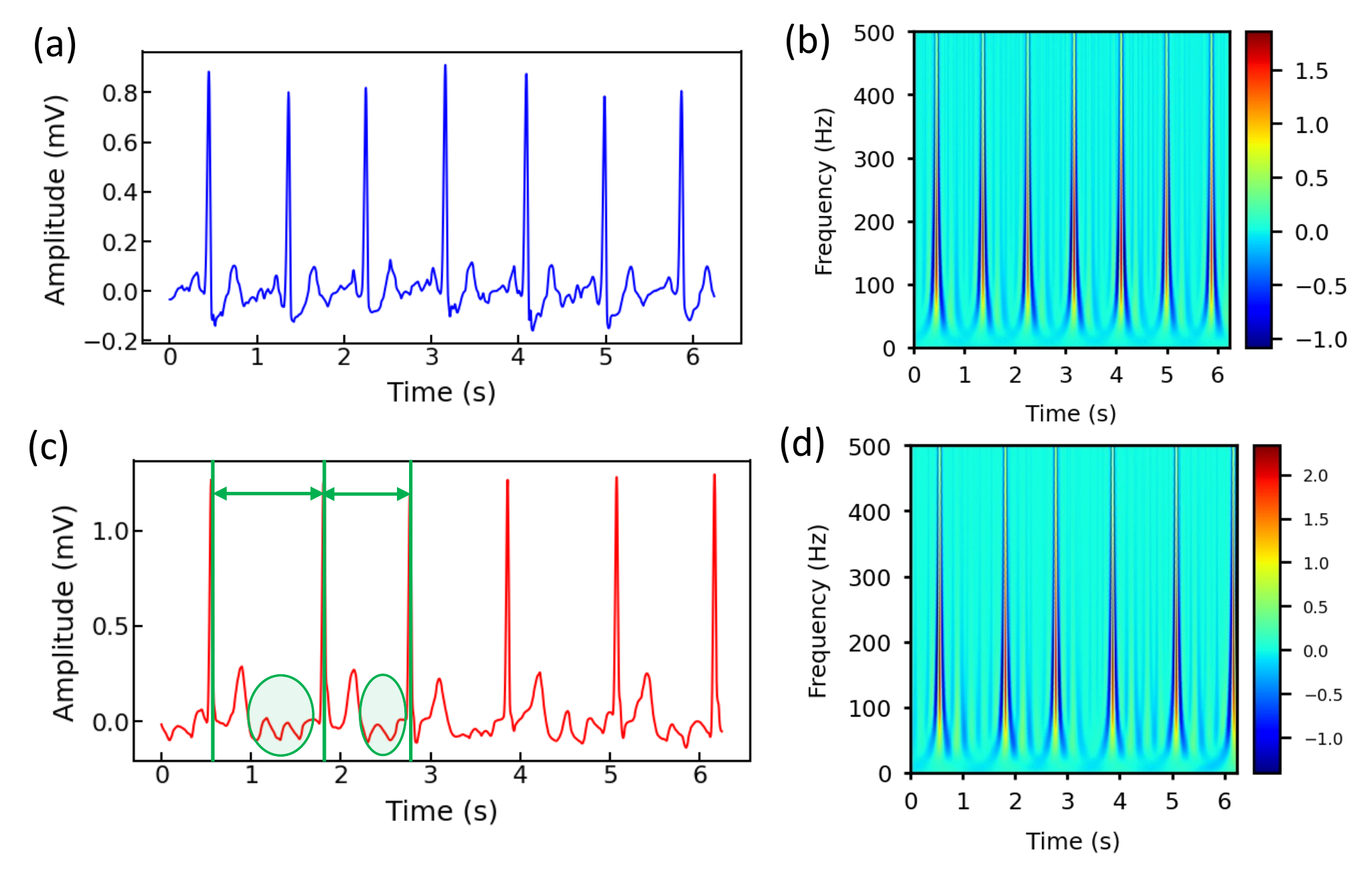}
	%\centerline{\includegraphics{fig1.png}}
	\caption{(a) The raw ECG signal from Physionet labeled as normal and (b) its corresponding 2D CWT scalogram. (c) The raw ECG signal from Physionet labeled as AF and (d) its corresponding 2D CWT scalogram. Note that the RR intervals are different in the AF sample and irregular F waves (circled) appear in (c).
	}
	\label{Fig:physioCWT}
\end{figure}

Fig. \ref{Fig:physioCWT} (a-b) and (c-d) show the healthy and AF examples of the raw ECG signals obtained from PhysioNet and their 2D time-frequency patterns after CWT transformation with mexh wavelet, respectively. The colors in the scalogram indicate the energy density of the signal component at the corresponding frequency and time \cite{addison2005wavelet, he2018automatic}. According to Fig. \ref{Fig:physioCWT} (a) and (c), two general differences can be observed: 1) 
The AF ECG signal lacks a distinct P wave, while it shows a fast and chaotic F wave due to the atrial fluttering (Fig. \ref{Fig:physioCWT} (c)), in comparison to a normal ECG signal (Fig. \ref{Fig:physioCWT} (a));
2) Irregular RR intervals are observed in AF ECG (Fig.\ref{Fig:physioCWT} (c)) caused by a non-synchronized ventricular response to the abnormal atrial excitation \cite{he2018automatic}. The discriminative information in the time domain can also be captured by the CWT scalograms shown in Fig. \ref{Fig:physioCWT}(b) and (d). By using a 2D CNN to analyze the visual representation of 2D time-frequency scalograms, we can better understand the features that distinguish AF from normal heart rhythms and make more accurate predictions.

\subsection{Convolutional Neural Network}
We engage CNN to build a data-driven classifier for differentiating AF samples from normal ECG samples. CNN is a type of neural network architecture specifically designed to process data that has a grid-like structure such as images \cite{khan2020survey}. As opposed to traditional multilayer perceptron networks (MLPs), where the input of each neuron consists of the outputs of all the neurons from the previous layer, the neuron in CNN only receives its input from a localized region of the previous layer, known as its receptive field. The main building blocks of a CNN are convolutional layers, pooling layers, and fully connected layers.

Convolutional layers are responsible for performing a convolution operation on the input data, using a set of filters to extract local features in the data, and producing a feature map that summarizes such local information. Let $\theta$ and $X$ denote the filter (also known as the kernel) and the input. The convolution operation works as follows:
 \begin{eqnarray}
 	(X\otimes \theta)_{ij}=\sum_{m=0}^{s_1-1}\sum_{n=0}^{s_2-1}X(i+m,j+n)\theta(m,n)
 \end{eqnarray}
where $s_1$ and $s_2$ denote the size of the 2D kernel, and $(i,j)$ denotes the location on the 2D input (e.g., image). After being applied with the activation function, the feature map of the input is obtained as \cite{lecun1995convolutional,jing2021ecg}:
\begin{equation}
	X_q^{l} = \sigma\left(\sum_{p} \theta_{pq}^l \otimes X_p^{l-1} +b_q^l\right)
\end{equation}
where $X_q^{l}$ is the $q$-th feature at layer $l$, $X_p^{l-1}$ is the $p$-th input feature map of the previous $(l-1)$-th layer, $\sigma$ denotes the activation function to induce the non-linearity in the functional mapping, and $b_q$ represents the bias. This procedure is repeated by applying multiple filters to generate an arbitrary number of feature maps to capture different characteristics of the input. Note that kernels are shared across all the input positions, which is also called weight sharing, the key feature of CNN. The weight-sharing technique guarantees the extracted local patterns are translation invariant and increases computational efficiency by reducing the model parameters to learn compared with fully connected neural networks.

The pooling layer mimics the human visual system by combining the outputs of multiple neurons (i.e., clusters) into a single neuron in the next layer, effectively creating a condensed representation of the input. The pooling significantly reduces the spatial resolution and only focuses on the prominent patterns of the feature maps, making the network more robust to small translations and distortion in the input data \cite{xia2018detecting}. Popular pooling techniques include maximum pooling, average pooling, stochastic pooling, and adaptive pooling, which are typically performed on the values in a sub-region of the feature map \cite{akhtar2020interpretation}.

The fully connected layers form a dense network that can learn complex non-linear relationships between the inputs and outputs. It takes the output of the previous layer, which is typically a high-dimensional tensor containing discriminant features extracted by convolutional and pooling layers, and flattens it into a one-dimensional vector. This vector is then used as the input to the fully-connected layer. The fully-connected layer is similar to an MLP in that every neuron in one layer is connected to every neuron in the next layer. By using a proper activation function, the neural network is able to produce classification decisions \cite{nurmaini2020robust}. By stacking these building blocks (convolutional layers, pooling layers, and fully connected layers) in various combinations, CNN is able to learn complex features in the input data, allowing them to effectively solve a wide range of image and signal processing tasks \cite{andreotti2017comparing}.

\subsubsection{2D CNN with ResNet}

We propose to engage 2D CNN to investigate the 2D time-frequency scalograms converted from denoised ECG signals by CWT for AF identification. It has been demonstrated that the substantial depth of the convolutional network is beneficial to the network performance \cite{simonyan2014very} . % and a deeper CNN should {\color{red}yield no higher training error than a CNN with fewer layers (confusing, no higher training error doesn't mean better)}. 
{However, as the number of convolutional layers increases, the training loss stops further decreasing and becomes saturated because of the gradient dissipation issue. As such, a CNN with a deeper architecture, counterintuitively, sometimes incurs a larger training error compared to its shallow counterpart upon convergence \cite{he2016deep}. In order to solve such network degradation and gradient vanishing problems, the residual network (ResNet) has been developed to improve the accuracy of CNNs with considerably increased depth. 

The core of ResNet is the residual learning technique \cite{he2016deep}. Specifically, instead of using the stacked convolutional layers to directly fit the underlying mapping from the input to the output, ResNet focuses on fitting a residual mapping. Fig. \ref{Fig:block} shows a ResNet building block with input $X$ and its corresponding output mapping $Y$. The residual block engages a shortcut connection that bypasses one or more convolutional layers and allows the information to flow directly from the input to the output. As such, the input $X$ is added to the output of the block $F(X)$ (enclosed by the dashed circle in Fig. \ref{Fig:block}, allowing the network to learn the residual mapping represented as $Y=F(X)+X$ instead of learning the direct mapping as $Y=F(X)$. This design mitigates the gradient vanishing problem and allows for deeper networks to be trained effectively.

In our study, we engage the ResNet with 18 layers (ResNet18) to build the AF classifier because ResNet18 has been proven to be able to generate a comparable result with a faster convergence compared to a deeper counterpart  \cite{he2016deep}. Fig. \ref{Fig:resnet} shows the detailed structure of ResNet18. Note that the notation of $\text{2DConv}(n_{input}, n_{output}, n_{fdim1}\times n_{fdim2})$ denotes that, in the current 2D convolutional layer, there are $n_{input}$ input channels, $n_{output}$ output channels (i.e., number of filters) with the 2D filter size of $n_{fdim1} \times n_{fdim2}$. For example, $(64, 128, 3\times 3)$ indicates that this convolutional layer is composed of 128 filters with the filter size of $3\times 3$ applied on the input data with 64 channels. 

\begin{figure}
	\centering
	\includegraphics[width=3in]{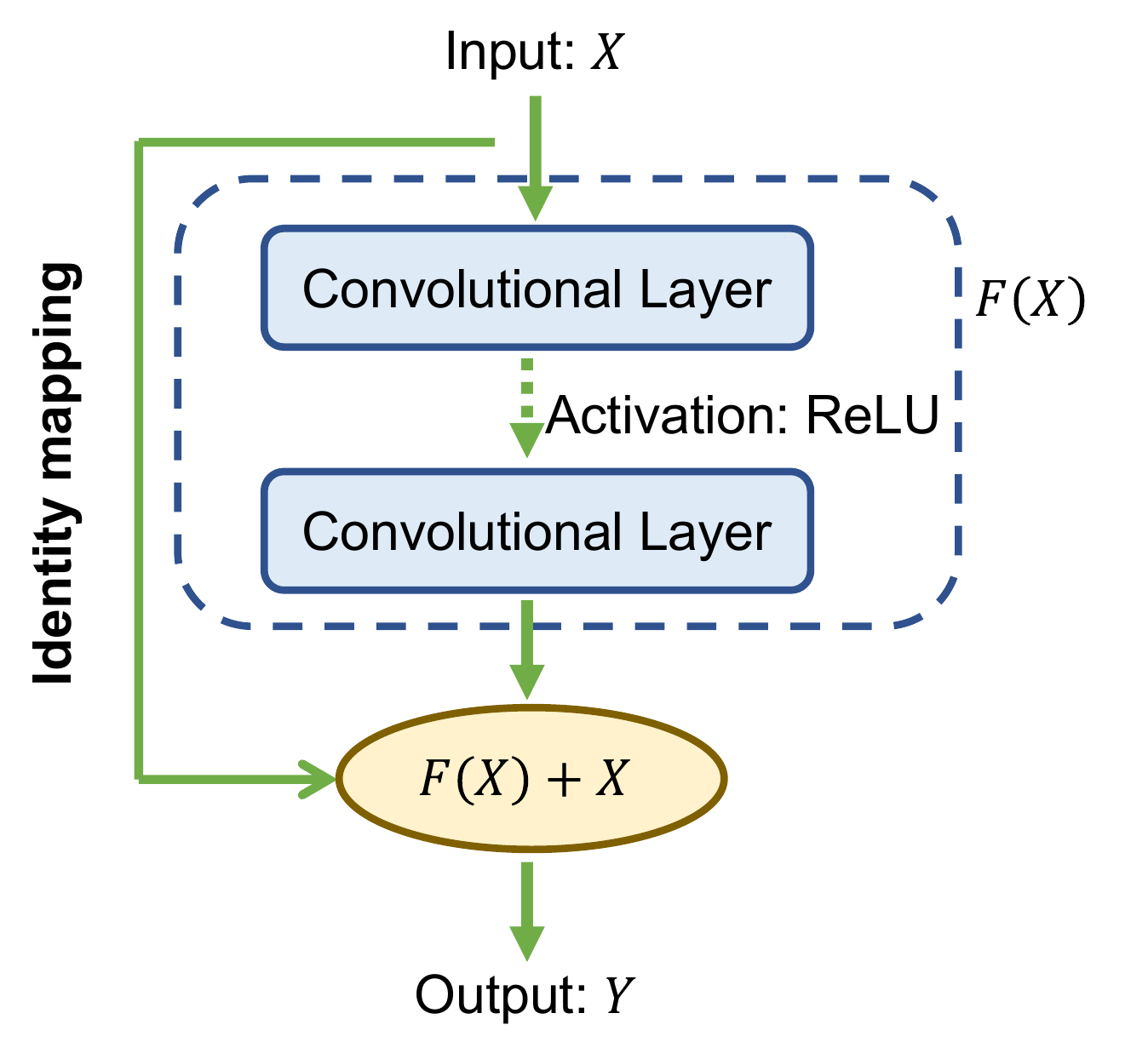}
	%\centerline{\includegraphics{fig1.png}}
	\caption{A building block of the ResNet. 
	}
	\label{Fig:block}
\end{figure}

\begin{figure}
	\centering
	\includegraphics[width=4.0in]{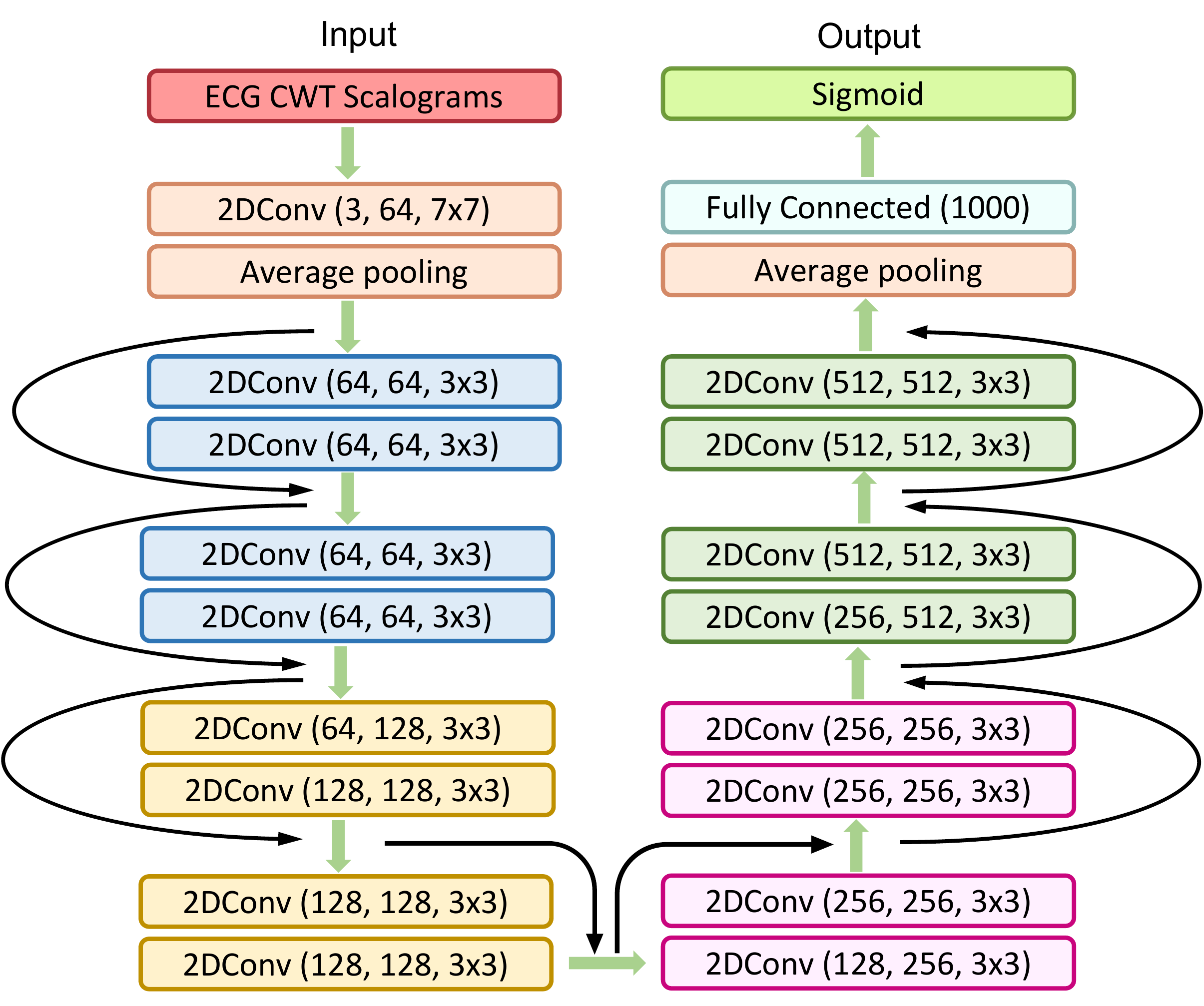}
	%\centerline{\includegraphics{fig1.png}}
	\caption{The detailed architecture of ResNet18.
	}
	\label{Fig:resnet}
\end{figure}

\subsubsection{Multi-Branching Convolutional network}
Data-driven identification of AF from ECG recordings generally suffers from imbalanced data issues. The obtained ECG signals contain far more normal samples than AF in both the PhysioNet and OUHSC datasets, with imbalanced ratios of approximately 7:1. Here, we propose to incorporate a multi-branching (MB) technique into ResNet18 (MB-ResNet) \cite{wang2021multi} to relieve the imbalanced data issue and enhance the performance of the classifiers. The schematic of MB inputs and outputs is shown in Fig. \ref{Fig:branch}. Specifically, $N_b$ balanced datasets are passed through a core ResNet and generate $N_b$ branching outputs. Note that $N_b$ balanced datasets are drawn from the original data $D=\{D_-, D_+\}$, by undersampling the majority normal ECG samples $D_-$ and formulate a balanced dataset as $D_i = \{D_-^i, D_+\}$, where $D_+$ stands for the minority class, i.e., the entire AF training samples, and $D_- = \cup_{i\in 1,\dots,N_b}D_-^i$. $N_b$ outputs will be generated with respect to its corresponding balanced input. The core ResNet is trained by all the $N_b$ datasets, whereas each output branch is trained by the corresponding balanced input sub-dataset.

In the current investigation, we aim to identify AF samples from normal ECG samples. The neural network is expected to produce high probabilities (close to 1) for AF samples and low probabilities (close to 0) for normal ECG samples. We choose the binary cross-entropy as the loss function for MB-ResNet, which is defined as:
\begin{equation}
	\begin{aligned}
		\mathcal{L}(\boldsymbol{\omega} ; D)= & -\sum_{j=1}^{N_d} \sum_{i=1}^{N_b} \mathcal{I}\left(j \in D_i\right)\left(y_j \log \left(\hat{P}_i\left(\boldsymbol{\omega} ; X^j\right)\right)\right. \left.+\left(1-y_j\right) \log \left(1-\hat{P}_i\left(\boldsymbol{\omega} ; X^j\right)\right)\right)
	\end{aligned}
\end{equation}
where $\vect{\omega}$ denotes the neural network parameter, $X^j$ and $y_j$ stand for one input sample and its corresponding true label respectively, $\mathcal{I}(\cdot)$ denotes the indicator function, $N_d$ is the total number of the training samples, and $\hat{P}_i\left(\boldsymbol{\omega} ; X^j\right)$ represents the predicted probability for AF at the $i$-th branching output given the input signal $X^j$. 

The adaptive momentum method (Adam) \cite{kingma2014adam} is adopted to minimize the loss function and update the neural network parameters. %For each iteration, we use the entire minority dataset (AF samples) $D_+$ and a subset of majority class (normal samples) $D_-^i$ with the same size of $D_+$ for training. Then, we randomly choose one branching dataset to train the network and optimize the neural network parameters $\vect{\theta}$. 
In the inference stage, the MB network generates $N_b$ predictions for AF probability, which correspond to the $N_b$ branching outputs. The final predicted probability for AF ($\hat{P}$) is determined by taking the average of the $N_b$ outputs:
$$\hat{P} = \frac{1}{N_b}\sum_{i=1}^{N_b}\hat{P}_i $$
where $\hat{P}_i$ is the predicted probability of $i$-th branching output. 

\begin{figure}
	\centering
	\includegraphics[width=2in]{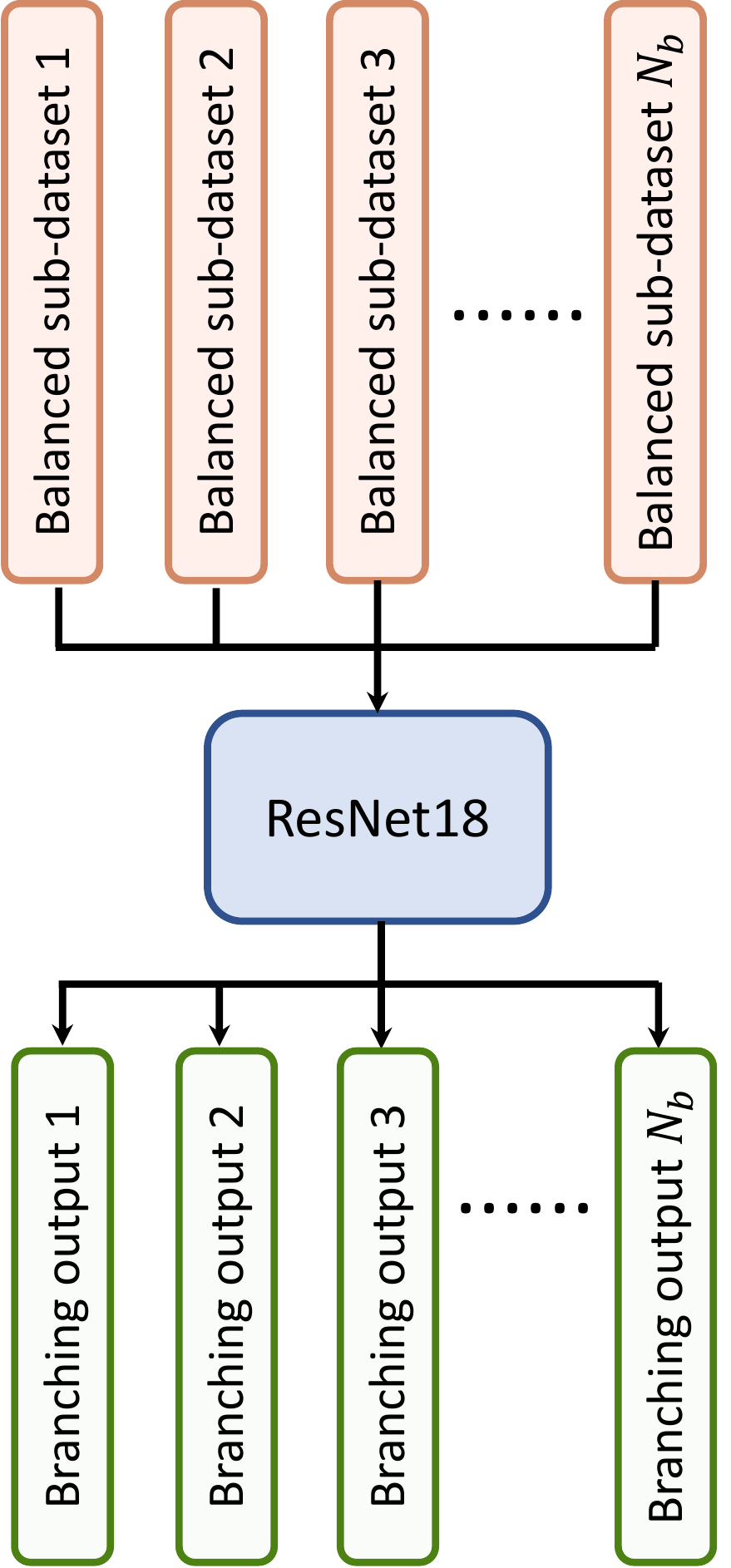}
	%\centerline{\includegraphics{fig1.png}}
	\caption{Illustration of the multi-branching architecture. 
	}
	\label{Fig:branch}
\end{figure}

\section{Experimental Design and Results } 
\label{s:results}

 \subsection{Experimental Design}

We validate and evaluate the performance of the proposed CWT-MB-ResNet framework using both Physionet Challenge and OUHSC datasets. The experiment design is shown in Fig. \ref{Fig:flowchart}. We compare the performance of our CWT-MB-ResNet with 1D-CNN (Fig. \ref{Fig:flowchart} (a)), 1D-CNN with the multi-branching network (1D-MB-CNN) (Fig. \ref{Fig:flowchart} (b)), and ResNet with CWT features (CWT-ResNet). Note that the input of 1D-CNN and 1D-MB-CNN consists of the denoised ECG time series. The detailed 1D-CNN architecture is illustrated in Fig. \ref{Fig:1D-CNN}, including three convolutional layers followed by pooling layers to reduce the dimensionality of the data, a batch-normalization layer to stabilize the network training, and one fully connected layer to make the final prediction. Note that the notation of $\text{1DConv}(n_{input}, n_{output}, n_{fdim})$ indicates that, in the current 1D convolutional layer, there are $n_{input}$ input channels and $n_{output}$ output channels (i.e., number of filters) with a 1D filter size of $n_{fdim}$.

\begin{figure}
	\centering
	\includegraphics[width=5in]{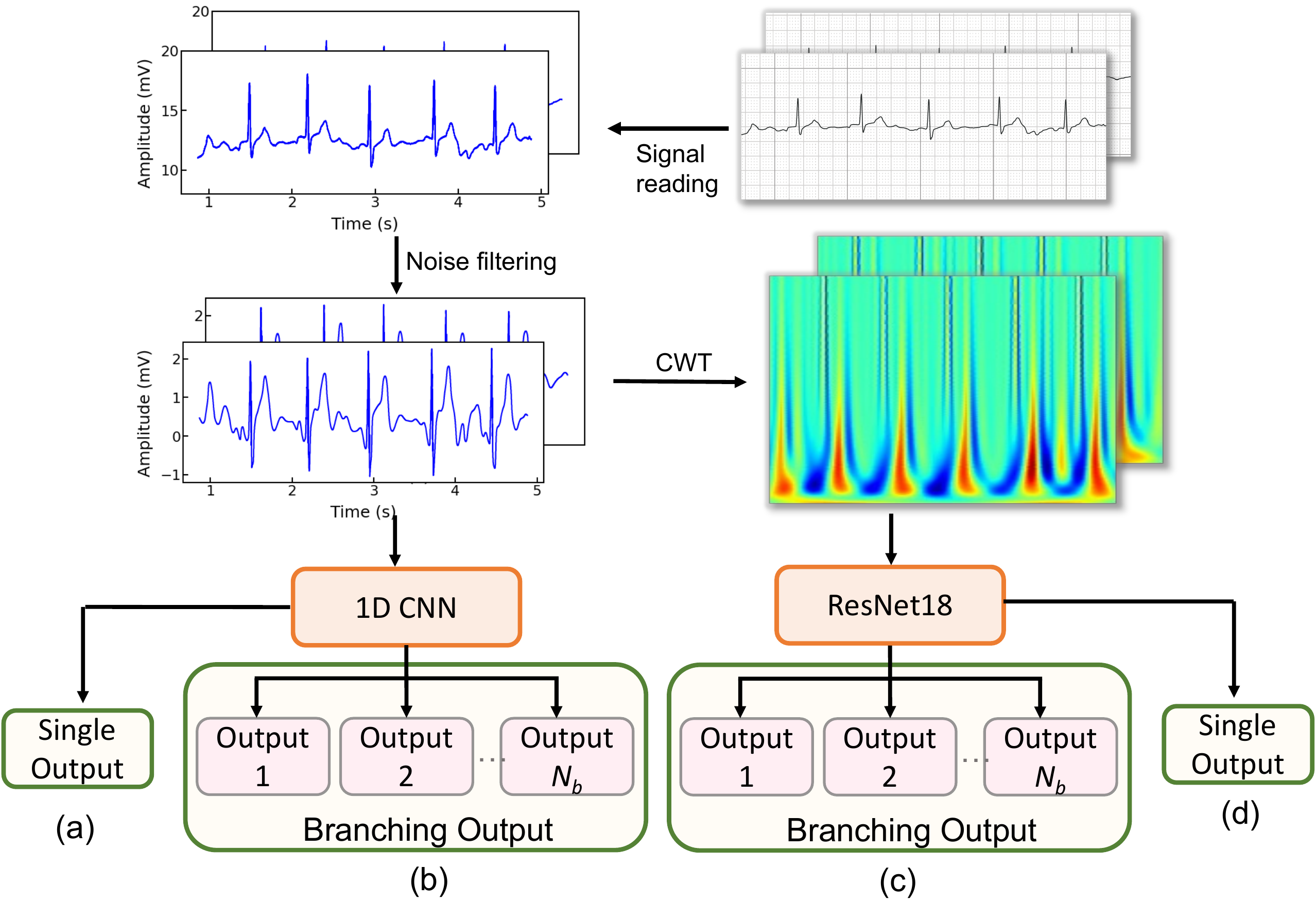}
	%\centerline{\includegraphics{fig1.png}}
	\caption{The flowchart of the experimental design: (a) 1D-CNN; (b) 1D-MB-CNN; (c) CWT-MB-ResNet; (d) CWT-ResNet. 
	}
	\label{Fig:flowchart}
\end{figure}

 The classification performance will be evaluated with three metrics: Receiver-Operating-Characteristic Curve (ROC), Precision-Recall Curve (PRC), and F1 score, which will be calculated using the test set.  
The ROC provides the graphic representation of the trade-off between the true positive rate (TPR) and the false positive rate (FPR) for different threshold settings. The area under ROC (AUROC) is often used as a metric to compare different models, with a larger AUROC indicating a better-performing classifier. A good model typically has a ROC curve that is situated toward the top-left corner of the graph.
The PRC illustrates the interplay between a predictive model's precision and recall metrics across a range of probability thresholds.  A good classifier has the PR curve towards the top-right corner. A higher area under PRC (AUPRC) value suggests a more effective model. The F1 score quantifies the equilibrium between a model's precision and recall for a binary classifier system by computing their harmonic mean, which is defined as 
$$F1 = \frac{2*Precision*Recall}{Precision+Recall}$$
Note that the F1 score ranges from 0 to 1, where a score of 1 indicates a perfect balance between precision and recall.

\begin{figure}
	\centering
	\includegraphics[width=1.6in]{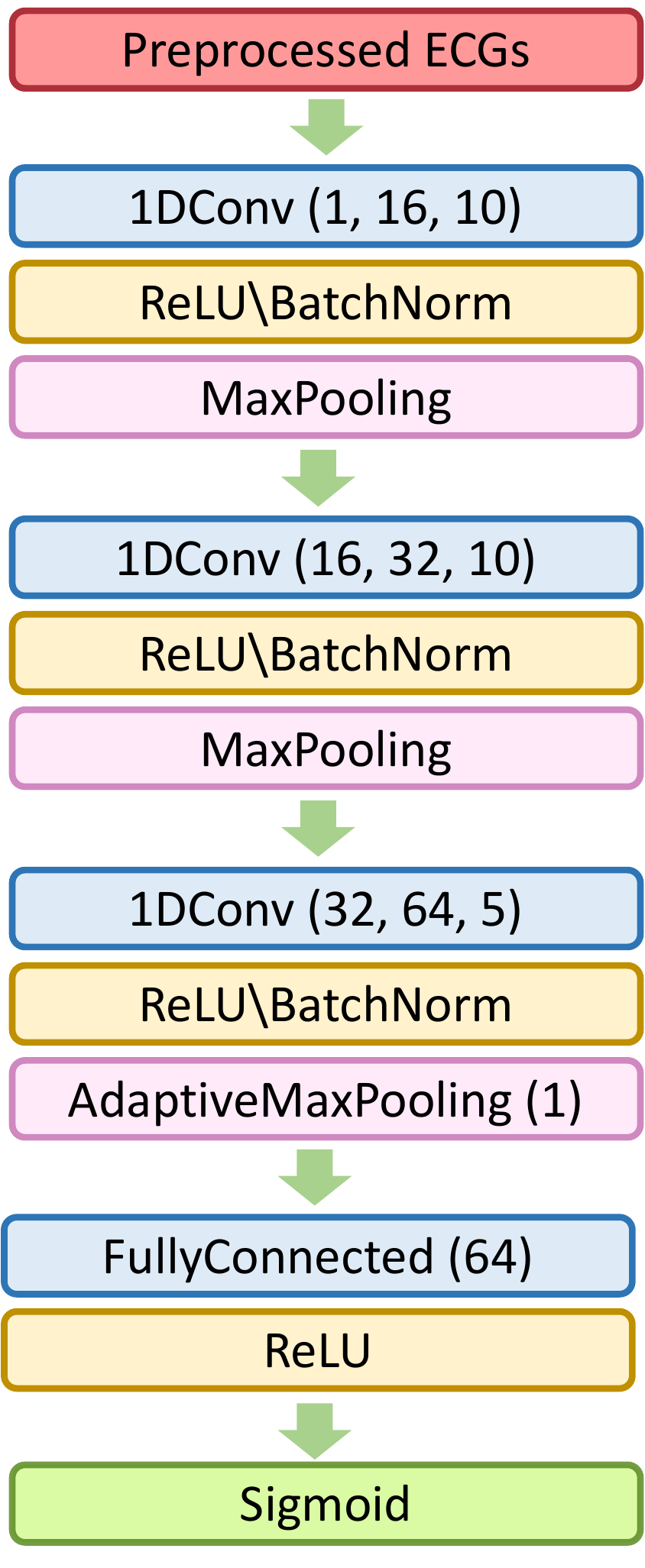}
	\caption{The 1D-CNN architecture.
	}
	\label{Fig:1D-CNN}
\end{figure}

\subsection{Experimental Results from the OUHSC Dataset}
 
Fig. \ref{Fig:aucpr_ou} displays the ROC and PR curves of all four models using the OUHSC dataset. 
The 2D ResNet models (i.e., CWT-ResNet and CWT-MB-ResNet), which use 2D scalograms transformed from ECG signals as the input, produce a larger area under the curves (both ROC and PRC) compared to their 1D counterparts (i.e., 1D-CNN and 1D-MB-CNN). This demonstrates the efficacy of using the CWT to extract time-frequency features in the ECG signal analysis. Additionally, the models with an MB architecture (i.e., 1D-MB-CNN and CWT-MB-ResNet) produce a larger area under both the ROC and PR curves compared to models without MB outputs (i.e., 1D-CNN and CWT-ResNet), which highlights the effectiveness of using the MB structure in addressing imbalanced data issues. The ROC and PR plots demonstrate the superiority and robustness of the proposed CWT-MB-ResNet framework for identifying the AF samples.
   \begin{figure}
   	\centering
   	\includegraphics[width=5in]{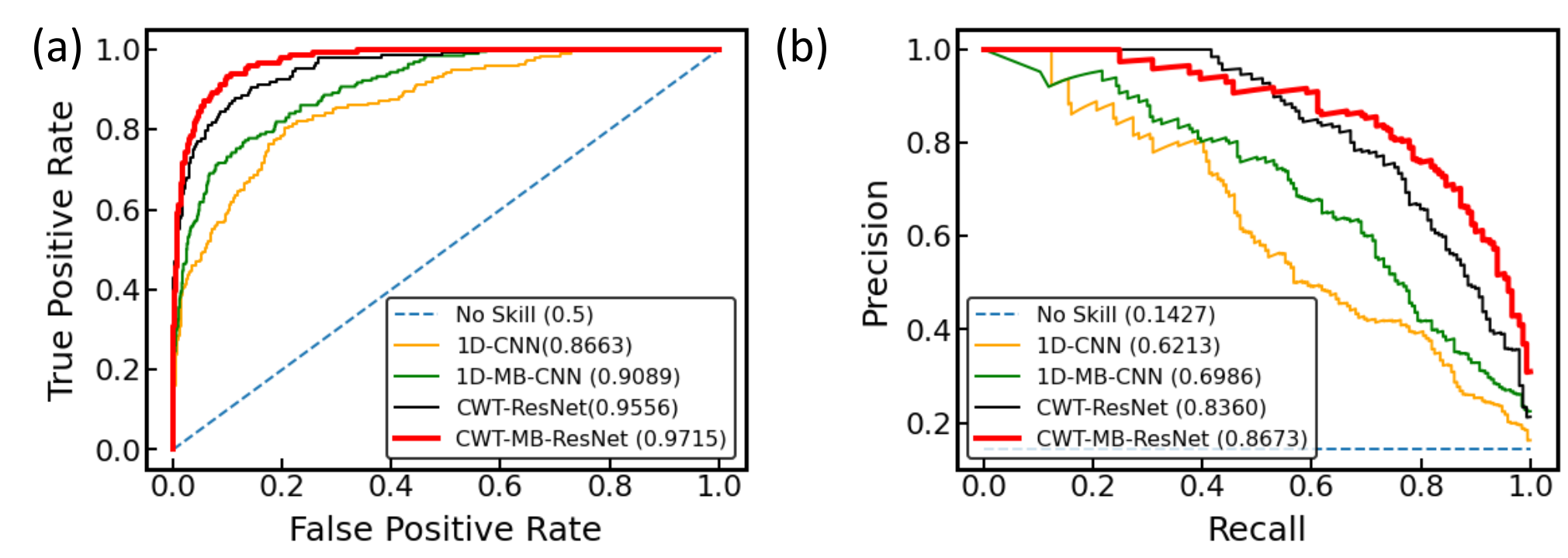}
   	\caption{The comparison of (a) ROC and (b) PRC among different models using the OUHSC data. 
   	}
   	\label{Fig:aucpr_ou}
   \end{figure}

Table. \ref{table:ou} shows AUROC, AUPRC, and F1 scores generated from the four methods using the OUHSC dataset. Our CWT-MB-ResNet method generates the best AUROC, AUPRC, and F1 scores with the values of 97.15\%, 86.73\%, and 0.8155. Note that the MB technique demonstrates its effectiveness on both 1D-CNN and CWT-ResNet as the AUROC, AUPRC, and F1 scores provided by the MB-based neural network models are higher than their non-MB-based counterparts. Moreover, the AF classifier using 2D-CNN-based ResNet18 supported by the time-frequency transformation of ECG time series presents a more potent predictive power than time sequence classification using 1D CNN. For example, CWT-MB-ResNet improves the  AUROC, AUPRC, and F1 scores from 90.89\%, 69.86\%, and 0.6601 to 97.15\%, 86.73\%, and 0.8155 respectively compared with the 1D-MB-CNN.

\subsection{Experimental Results from the Physionet/CinC 2017 Challenge Dataset}

Fig. \ref{Fig:aucpr_phy} further shows the ROC and PRC analysis for the Physionet/Cinc 2017 challenge dataset. Similar to the results from the OUHSC dataset, the 2D ResNet models (CWT-ResNet and CWT-MB-ResNet) with 2D scalograms of ECG signals as the input, outperform their 1D counterparts (1D-CNN and 1D-MB-CNN) in terms of the area under the ROC and PR curves. Furthermore, the MB-based models (1D-MB-CNN and CWT-MB-ResNet) effectively account for the imbalanced data issues, exhibiting better performance compared to the non-MB-based models (1D-CNN and CWT-ResNet). Table. \ref{table:phy} demonstrates the comparison of AUROC, AUPRC, and F1 scores provided by 1D-CNN, 1D-MB-CNN, CWT-ResNet, and CWT-MB-ResNet. Our CWT-MB-ResNet yields the best classification performance among the four methods, generating the highest AUROC, AUPRC, and F1 scores of 97.41\%, 93.53\%, and 0.8865. Especially, our CWT-MB-ResNet model improves the F1 score by 46.2\% percent compared to the pure 1D-CNN with no CWT transform or MB structure.

\begin{figure}
	\centering
	\includegraphics[width=5in]{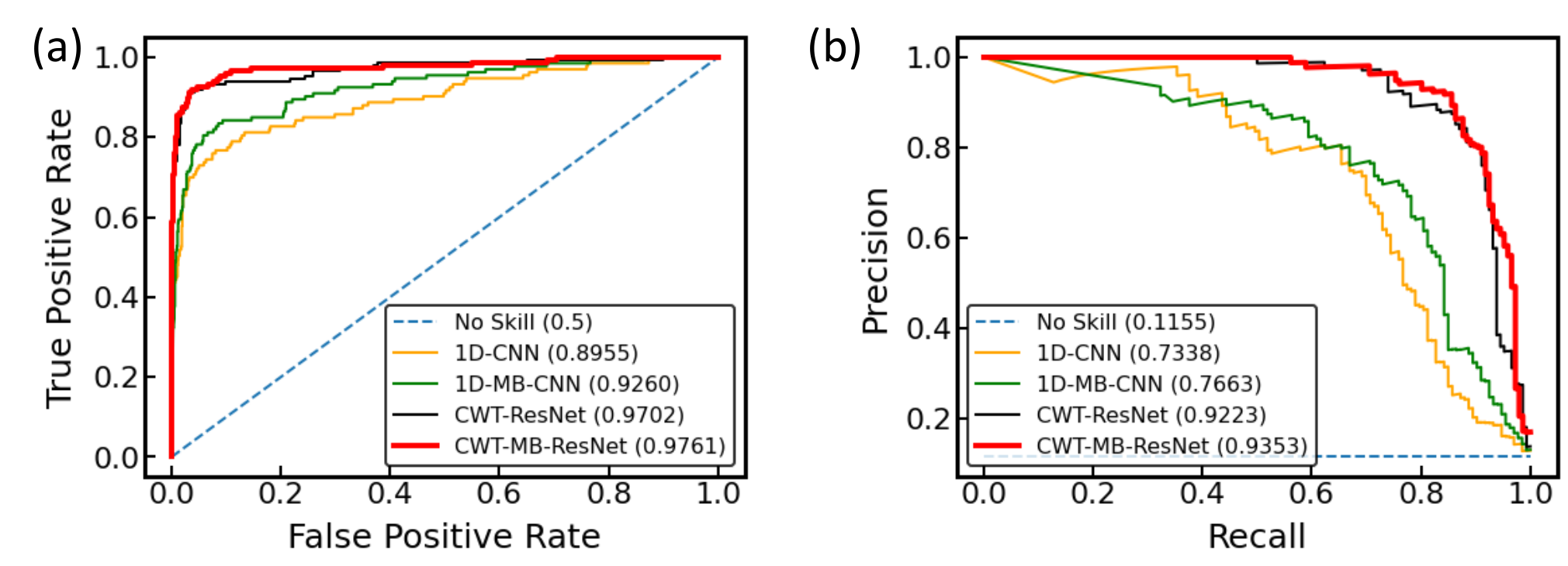}
	\caption{The comparison of (a) ROC and (b) PRC between different models using data from Physionet/Cinc 2017 challenge.
	}
	\label{Fig:aucpr_phy}
\end{figure}

\begin{table}
	
	\caption{The comparison of AUROC, AUPRC, and F1 scores generated from 1D-CNN, 1D-MB-CNN, CWT-ResNet, and CWT-MB-ResNet using OUHSC data. }
	\centering
	{\scriptsize
		\resizebox*{0.8\textwidth}{!}{%
			\begin{tabular}{ccccc}
				\toprule[1.0pt]
				\multirow{1}{*}{} & 1D-CNN & 1D-MB-CNN & CWT-ResNet & CWT-MB-ResNet \\
				\cmidrule(lr){1-5}
				\multirow{1}{*}{AUROC} & 86.63\% & 90.89\% & 95.56\% & 97.15\% \\
				\cmidrule(lr){1-5}
				\multirow{1}{*}{AUPRC} & 62.13\% & 69.86\% & 83.60\% & 86.73\% \\
				\cmidrule(lr){1-5}
				\multirow{1}{*}{F1} & 0.5573 & 0.6601 & 0.7542 & 0.8155 \\		
				\bottomrule[1.0pt]
			\end{tabular}
	}}
	\label{table:ou}
\end{table}

\begin{table}
	
	\caption{The comparison of AUROC, AUPRC, and F1 scores generated from 1D-CNN, 1D-MB-CNN, CWT-ResNet, and CWT-MB-ResNet using data from Physionet/CinC 2017 challenge.  }
	\centering
	{\scriptsize
		\resizebox*{0.8\textwidth}{!}{%
			\begin{tabular}{ccccc}
				\toprule[1.0pt]
				\multirow{1}{*}{} & 1D-CNN & 1D-MB-CNN & CWT-ResNet & CWT-MB-ResNet \\
				\cmidrule(lr){1-5}
				\multirow{1}{*}{AUROC} & 89.55\% & 92.60\% & 97.02 \% & 97.61\% \\
				\cmidrule(lr){1-5}
				\multirow{1}{*}{AUPRC} & 73.38\% & 76.63\% & 92.23\% & 93.53\% \\
				\cmidrule(lr){1-5}
				\multirow{1}{*}{F1} & 0.7219 & 0.7380 & 0.8690 & 0.8865 \\		
				\bottomrule[1.0pt]
			\end{tabular}
	}}
	\label{table:phy}
\end{table}

\begin{table}
	
	\caption{The comparison of F1 scores between the proposed CWT-MB-ResNet method with existing literature using data from Physionet/CinC 2017 challenge.  }
	\centering
	{\scriptsize
		\resizebox*{1\textwidth}{!}{%
			\begin{tabular}{ c|c|c } 
				\toprule[1.5pt]
				\multirow{1}{*}{Authors} & Methods  & F1\\
				\midrule[0.5pt]
				\multirow{1}{*}{\parbox{6cm}{\centering Andreotti \textit{et al.} \cite{andreotti2017comparing}}}  & {\parbox{3cm}{\centering ResNet}}  & {\parbox{2cm}{\centering 0.78}} \\
				\multirow{1}{*}{Chandra \textit{et al.} \cite{chandra2017atrial}} & CNN & 0.73 \\ 
				\multirow{1}{*}{ Ghiasi \textit{et al.} \cite{ghiasi2017atrial}} & Feature-based algorithm and CNN & 0.72 \\ 
				\multirow{1}{*}{  Schwab \textit{et al.} \cite{schwab2017beat}} & RNN & 0.79 \\ 
				\multirow{1}{*}{ Limam \textit{et al.} \cite{limam2017atrial}} & CRNN & 0.85 \\ 
				\multirow{1}{*}{  This paper  } & CWT-MB-ResNet & 0.8865 \\ 
				\bottomrule[1.5pt]		
			\end{tabular}
	}}
	\label{table:comp}
\end{table}

We further compare the proposed CWT-MB-ResNet model with existing studies in the literature that also used deep learning models for AF classification using the dataset from the PhysioNet/CinC challenge 2017. Table. \ref{table:comp} summarizes the comparison results in terms of F1 score. Our proposed CWT-MB-CNN demonstrates the best F1 score compared with the other framework settings and neural network structures. This is due to the fact that our CWT-MB-ResNet not only leverages CWT to capture comprehensive 2D time-frequency features from ECG signals but also incorporates the MB structure to effectively cope with the imbalanced data issues.

\section{Conclusions} \label{s:conclusion}
In this paper, we develop a novel framework based on Continous Wavelet Transform (CWT) and multi-branching ResNet for AF identification. We first transform the 1D ECG time series into 2D time-frequency scalograms to avoid the aliasing of multiple frequency components, which can serve as the input to the 2D CNN-based classifier. Second, we leverage the ResNet architecture to cope with the possible gradient dissipation problems in deep 2D CNN and increase the effectiveness of network training. Moreover, a multi-branching architecture is incorporated into the ResNet to mitigate the possible prediction bias caused by the imbalanced data issue. Finally, we implement the proposed CWT-MB-ResNet to predict AF using the ECG recordings from PhysioNet/CinC Challenge 2017 and the ECG PDFs from OUHSC. Experimental results show that the proposed CWT-MB-ResNet achieves the best prediction performance for both datasets in AF detection. The CWT-MB-ResNet framework has great potential to be applied in clinical practice to improve the accuracy in ECG-based diagnosis of heart disease.

\bibliographystyle{IEEEtran}

\bibliography{refs}
	
\end{document}